# Autonomous Discovery of Unknown Reaction Pathways from Data by Chemical Reaction Neural Network

Weiqi Ji, Sili Deng[*]

Department of Mechanical Engineering, Massachusetts Institute of Technology, Cambridge, MA 02139.

Corresponding Author: * E-mail: silideng@mit.edu (Sili Deng)

ORCID: Weiqi Ji: 0000-0002-7097-0219, Sili Deng: 0000-0002-3421-7414.

**Abstract:** Chemical reactions occur in energy, environmental , biological, and many other natural systems, and the inference of the reaction networks is essential to understand and design the chemical processes in engineering and life sciences. Yet, revealing the reaction pathways for complex systems and processes is still challenging due to the lack of knowledge of the involved species and reactions. Here, we present a neural network approach that autonomously discovers reaction pathways from the time-resolved species concentration data. The proposed Chemical Reaction Neural Network (CRNN), by design, satisfies the fundamental physics laws, including the Law of Mass Action and the Arrhenius Law. Consequently, the CRNN is physically interpretable such that the reaction pathways can be interpreted, and the kinetic parameters can be quantified simultaneously from the weights of the neural network. The inference of the chemical pathways is accomplished by training the CRNN with species concentration data via stochastic gradient descent. We demonstrate the successful implementations and the robustness of the approach in elucidating the chemical reaction pathways of several chemical engineering and biochemical systems. The autonomous inference by the CRNN approach precludes the need for expert knowledge in proposing candidate networks and addresses the curse of dimensionality in complex systems. The physical interpretability also makes the CRNN capable of not only fitting the data for a given system but also developing knowledge of unknown pathways that could be generalized to similar chemical systems.

## 1. Introduction

Discovering the reaction network in the chemical processes in energy conversion, environmental engineering, and biology is paramount to understanding the mechanisms of pollution and disease, as well as developing mitigation strategies and drugs. The traditional approach to building chemical reaction models is based on ab initio calculations and reaction templates developed with expert knowledge [1]. In many complex reacting systems, however, ab initio calculation is computationally intractable, and limited prior knowledge of the reaction templates has been established. Instead, time-resolved species concentration data is available due to emerging sensing and measurement technologies [2,3] (e.g., from high-throughput experiments in biology and materials sciences). As a result, a new paradigm of data-driven modeling has been the focus of recent efforts [4–14]. While many statistical regression methods have successfully learned the kinetic parameters for given reaction pathways [15–19], further investigation is needed to simultaneously infer the reaction pathways and quantify the kinetic parameters from data.

Achieving both the interpretability and accuracy of the data-driven chemical models are important and challenging. Numerous recent approaches [7,10,12,13] leverage symbolic regression and sparse regression to identify the reaction pathways from the time series data of the species concentrations. Symbolic and sparse regression can produce physically interpretable kinetic models with known candidate pathways. However, such candidate pathways can only be proposed for a few relatively simple systems due to the curse of dimensionality, since the possible interactions among species increase dramatically with the number of species [20]. For example, the number of possible reaction pathways scales with the fourth power of the number of species if we only consider possible reactions involving two species in the reactants and two species in the products. In addition, multiple channel reaction pathways would require duplicated reactions, such as two or more reactions with the same set of reactants and products but different activation energies. On the contrary, neural networks have shown promise in autonomously learning features of interactions among high-dimensional inputs [21]. Traditional neural networks can approximate unknown reaction pathways [22], but the weights are difficult to interpret physically, i.e., interpreting the reaction pathways and rate constants from the neural network weights to be correlated to traditional chemical models, limiting the capability of model generalization. The booming of deep learning has greatly benefitted from designing problem-specific structures for neural network models. For example, the Convolutional Neural Network incorporates the scale and rotation invariant of the visions. By connecting the convolutional kernel with traditional image filers, the weights of the kernel and the features learned by the kernel can be partially interpreted [21]. Long et al. [23] have connected the residual neural networks to the first-order Euler method in the time-stepping numerical method and design specific convolutional kernels to reveal the differential operators and to discover governing partial differential equations from data. Conceptually inspired by these studies, we aim at designing neural networks with a problem-specific structure to learn chemical kinetic models from data.

In this work, we propose a Chemical Reaction Neural Network (CRNN) to identify reaction pathways from data without any prior knowledge of the chemical system. We enable the interpretability of the neural networks by encoding governing physics laws into the architecture of the neural network while leveraging the stochastic gradient descent that is widely utilized by the deep learning community to optimize high-dimensional parameters. The physically interpretable feature enables us to relate the weights of the neural network to reaction pathways as well as kinetic parameters and provide knowledge and insights on the chemical network. The capability of the autonomous discovery of unknown pathways enables us to learn from the given system and formulate reaction templates that can be generalized to similar systems, such that knowledge transfer can be achieved. Therefore, the CRNN approach enables autonomous inference of new chemical reaction systems without prior knowledge of the reaction templates and has the potential to transform our understanding of complex reacting systems.

## 2. Methods

Without loss of generality, we first derive the CRNN to represent an elementary reaction involving four species of [A, B, C, D] with the stoichiometric coefficients of $[v_A, v_B, v_C, v_D]$.

$$v_A \text{ A } + v_B \text{ B } => v_C \text{ C } + v_D \text{ D} \tag{1}$$

Recall the Law of Mass Action discovered by Guldberg in 1879, the reaction rate $r$ of Eq. **1** can be described by the power-law expression with the rate constant $k$ and species concentrations of the reactants $[A]$ and $[B]$, shown in Eq. **2**. The expression can also be written as a cascade of exponential operation and weighted summation of species concentrations in the logarithmic scale, shown in Eq. **3**.

$$r = k[A]^{v_A}[B]^{v_B}[C]^0[D]^0 \tag{2}$$

$$r = \exp(\ln k + v_A \ln[A] + v_B \ln[B] + 0 \ln[C] + 0 \ln[D]) \tag{3}$$

Recall the formula of a neuron $y = \sigma(\mathbf{w}\mathbf{x} + b)$, in which $\mathbf{x}$ is the input to the neuron, $y$ is the output, $\mathbf{w}$ are the weights, $b$ is the bias and $\sigma(\cdot)$ is the nonlinear activation function. Therefore, an elementary reaction can be represented as a neuron, as shown in Fig. **1a**. The inputs are the species concentrations in the logarithmic scale, and the outputs are the production rates of all species $\left[\frac{d[A]}{dt}, \frac{d[B]}{dt}, \frac{d[C]}{dt}, \frac{d[D]}{dt}\right]$, in short, $[[\dot{A}], [\dot{B}], [\dot{C}], [\dot{D}]]$. The weights in the input layer correspond to the reaction orders, i.e., $[v_A, v_B, 0, 0]$ for $[A, B, C, D]$, respectively, since any species not presented in the reactants have the weights of zero. The bias corresponds to the rate constant in the logarithmic scale. The weights in the output layer correspond to the stoichiometric coefficients, i.e., $[-v_A, -v_B, v_C, v_D]$ for $[A, B, C, D]$, respectively.

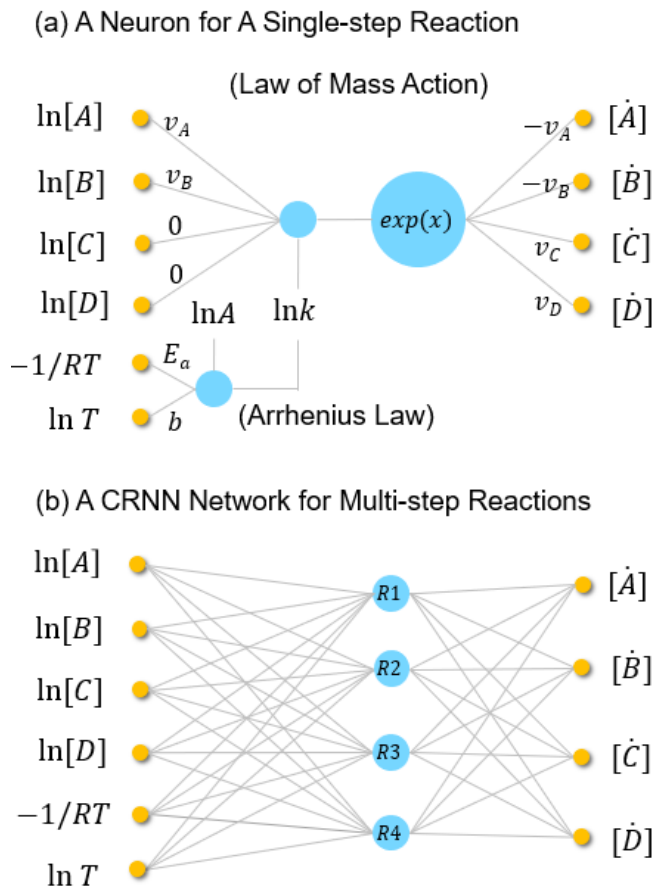

**Figure 1.** Schematic of the Chemical Reaction Neural Network illustrated for a reaction system with four species and four reaction steps. (a) First, a neuron is designed based on the Law of Mass Action and the Arrhenius Law, such that the learned neural network can be translated into interpretable reactions corresponding to the traditional chemical reaction network. (b) Then, the neurons are stacked into one hidden layer to formulate a CRNN for multi-step reactions.

In many chemical reaction systems, the rate constants are temperature dependent. For example, the Arrhenius Law discovered in 1889 can describe such dependence. The modified three-parameter Arrhenius formula states that

$$k = AT^b \exp\left(-\frac{E_a}{RT}\right),\tag{4}$$

where $A$ is the pre-factor or collision frequency factor, $b$ is a fitting parameter to capture the non-exponential temperature dependence, $E_a$ is the activation function, and $R$ is the gas constant. Equation **4** can also be written as a linear operation of temperature and the pre-factor A as shown in Eq. **5**, such that the Arrhenius Law is also represented in the CRNN, as shown in Fig. **1a**.

$$\ln k = \ln A + b \ln T - \frac{E_a}{RT}\tag{5}$$

A reaction network involving multiple elementary reactions is therefore represented by forming a neural network with one hidden layer. For example, the CRNN representation of a multi-step reaction network consisting of 4 species and 4 reactions is shown in Fig. **1b**. The number of hidden nodes is equal to the number of reactions.

Since the weights and biases in the CRNN are physically interpretable, the CRNN is essentially the digital twin of the classical chemical reaction network. The inference of the reaction network can then be accomplished by training the CRNN with experimental measurements. Considering a general chemical reaction system where the vector of species concentration $\boldsymbol{Y}$ evolves with time, we are trying to discover a CRNN that satisfies

$$\dot{\boldsymbol{Y}} = CRNN(\boldsymbol{Y}).$$

(6)

The CRNN can be trained from the concentration and production rate data pair of $\{\boldsymbol{Y}, \; \dot{\boldsymbol{Y}}\}$. In practice, we are able to measure the concentration time series data, while it is challenging to measure the derivative directly due to the presence of noise in the measurements. The production rate can be approximated from noisy data as demonstrated in a previous study [7,11]; however, additional efforts are required to tune the hyper-parameters, such as the regularization parameters, involved in the estimation [23]. Instead, we learn CRNN in the context of neural ordinary differential equations [24]. Specifically, an Ordinary Differential Equations (ODE) system can be formulated with Eq. **6** and numerically solved in an ODE integrator by providing initial conditions. The solution is denoted as $\boldsymbol{Y}^{CRNN}(t)$ in Eq. **7**, in which $\boldsymbol{Y_0}$ is a vector containing the initial conditions. We then define the loss function as the difference between the measured and predicted concentration time series data, as illustrated in Fig. **2a**. In the present work, the mean absolute error (MAE) is utilized as the loss metric in Eq. **8**.

$$\boldsymbol{Y}^{CRNN}(t) = \text{ODESolve}(CRNN(\boldsymbol{Y}), \boldsymbol{Y_0})$$

(7)

$$\text{Loss} = \text{MAE}(\boldsymbol{Y}^{CRNN}(t), \boldsymbol{Y}^{data}(t))$$

(8)

The recently developed differential programming package of DifferentialEquations.jl [25] written in the Julia language has readily enabled the computation of the gradients of the above loss functions with respect to the CRNN parameters via backpropagation over the ODE integrators. Then, we can use stochastic gradient descent optimization approach to learn the CRNN parameters [24], such as the popular optimizer of Adam proposed by Kingma and Ba [26]. We denote the framework of learning CRNN using neural ordinary differential equation as CRNN-ODE.

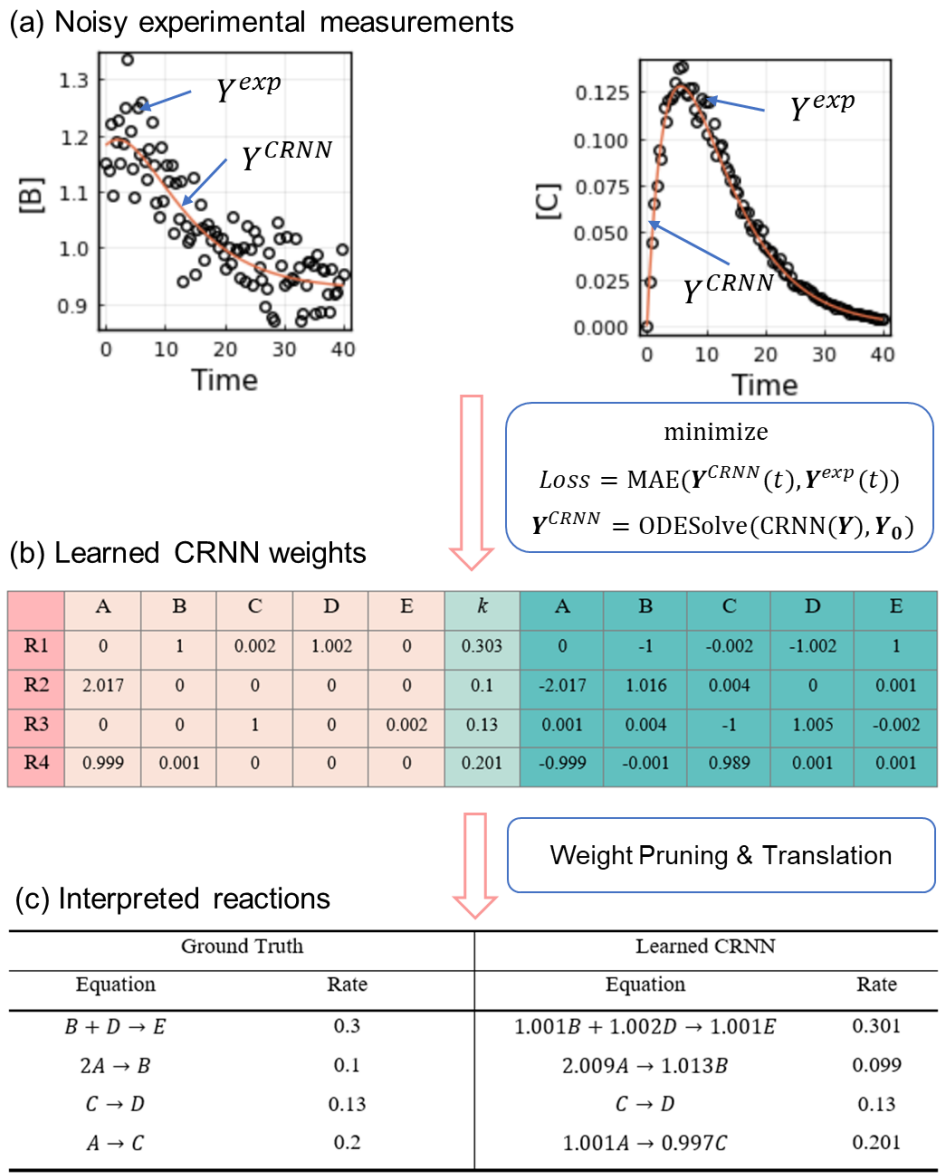

**Figure 2.** Schematic of the CRNN-ODE illustrated for a reaction system with five species [*A, B, C, D, E*] and four reaction steps. (a) $Y^{exp}$ corresponds to the noisy concentration time series, and $Y^{CRNN}$ refers to the solution of ODE integrations of the CRNN model. (b) The learned CRNN weights consist of three blocks: input weights as the reaction orders (left), biases as the rate constants (middle), and output weights as the stoichiometric coefficients (right). (c) The interpreted reactions translated from the pruned CRNN weights and biases.

With all of the species known, the number of nodes in the hidden layer is the major hyperparameter to be determined, which corresponds to the number of reactions involved in the CRNN. We propose a grid searching approach to determine the number of hidden nodes, i.e., increasing the number of proposed reactions until the performance cannot be further improved.

Finally, we employ hard threshold pruning to further encourage sparsity in the learned CRNN weights, especially the reaction orders and stoichiometric coefficients. The pruning proceeds by clipping the input and output weights below a certain threshold. The threshold is also determined by grid searching as will be illustrated in Section 3. Then we can translate the pruned CRNN model into the classical form of reaction equations. The effect of weight pruning is illustrated in Figs. **2b** and **2c**. Figure **2b** shows the learned CRNN weights without pruning and there exist many relatively small weights, such as the reaction order of 0.002 for species C in reaction R1. Those small weights are then pruned, and the interpreted reactions are presented in Fig. **2c**.

## 3. Results

We demonstrate the training of CRNN-ODE to simultaneously discover reaction pathways and learn kinetic parameters from synthetic noisy data in canonical chemical engineering and biochemistry systems, including both complete measurements and incomplete measurements with missing species. The ODE solver and the optimization are implemented with Julia and the code is open source. The code is available on GitHub at https://github.com/DENG-MIT/CRNN.

### 3.1 Case I: An Elementary Reaction Network Without Temperature Dependence

The first representative reaction network, taken from Searson et al. [27], comprises five chemical species labeled as $[A, B, C, D, E]$ that are involved in four reactions shown in Eq. **9** and Table **1**. Since the rate constants do not have temperature dependence, only the Law of Mass Action needs to be satisfied. The case is chosen to assess the capability of CRNN in discovering the chemical reaction pathways from noisy measurements and to demonstrate the CRNN-ODE algorithm.

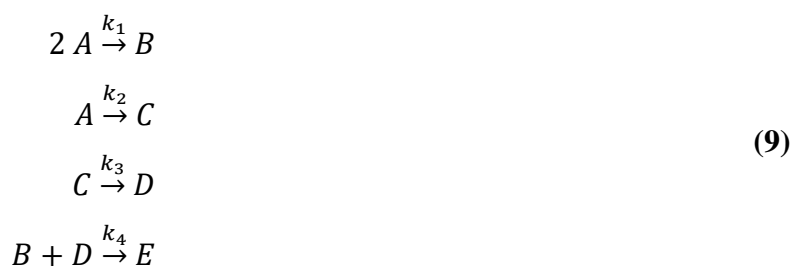

$$
\begin{aligned}
2\,A &\xrightarrow{k_1} B \\
A &\xrightarrow{k_2} C \\
C &\xrightarrow{k_3} D \\
B + D &\xrightarrow{k_4} E
\end{aligned}
\tag{9}
$$

A total of 30 synthetic experimental datasets are simulated with initial conditions randomly sampled between [0.2, 0.2, 0, 0, 0] and [1.2, 1.2, 0, 0, 0]. Each dataset comprises 100 data points evenly distributed in the temporal space with a timestep of 0.4 time unit. The simulation proceeds by solving the governing ordinary differential equations (shown in Supporting Information Eq. **S1**) using the solver of Tsitouras 5/4 Runge-Kutta method [25]. Gaussian noise is added to the species concentration profiles with the standard deviation of the noise being 5% of the concentrations.

**Table 1.** Learned pathways interpreted from the CRNN for Case I trained with noisy data with the standard deviation of the noise being 5% of the concentrations.

| Ground Truth | | Learned CRNN | |
|---|---|---|---|
| Equation | Rate | Equation | Rate |
| $B + D \rightarrow E$ | 0.3 | $B + 1.006D \rightarrow 1.006E$ | 0.307 |
| $2A \rightarrow B$ | 0.1 | $2.093A \rightarrow 1.107B$ | 0.101 |
| $A \rightarrow C$ | 0.2 | $1.004A \rightarrow 0.965C$ | 0.206 |
| $C \rightarrow D$ | 0.13 | $0.999C \rightarrow 1.011D$ | 0.13 |

The 30 datasets are randomly split into the training datasets and validation datasets by the ratio of 2:1, i.e., 20 of them as training datasets and 10 as validation datasets. The cross-validation help assess the level of over-fitting, and we adopt early stopping based on the loss in the validation dataset to prevent over-fitting [28]. The training of CRNN proceeds in a fashion of mini-batch. Specifically, for each parameter update, only one of the 20 training datasets is used for computing the loss function and the gradients to the CRNN model parameters. The mini-batch approach accelerates the training and implicitly regularizes the CRNN model [28]. The input weights share parameters with the output weights to accelerate optimization, i.e., the reaction orders are set to be equal to the stoichiometric coefficients for reactants. The optimizer of Adam [26] is adopted with a learning rate of 0.001.

We design a CRNN with five inputs to model case I. The first step of the CRNN-ODE modeling pipeline is to determine the number of hidden nodes. We then train CRNN with different numbers of hidden nodes to study the dependence of model performance on the number of hidden nodes. The model performance can be measured by the average MAE loss function across all experimental datasets. Figure **3** shows the typical evolution of loss functions with the number of epochs. Normally, the training converges at around 5000 epochs. Figure **4a** then shows the evolution of the minimum loss functions for both training and validation dataset with the number of hidden nodes. The training and validation loss decrease when the number of proposed reactions is less than four and reaches a plateau after that. It then can be inferred that the kinetics could be well described using four reactions.

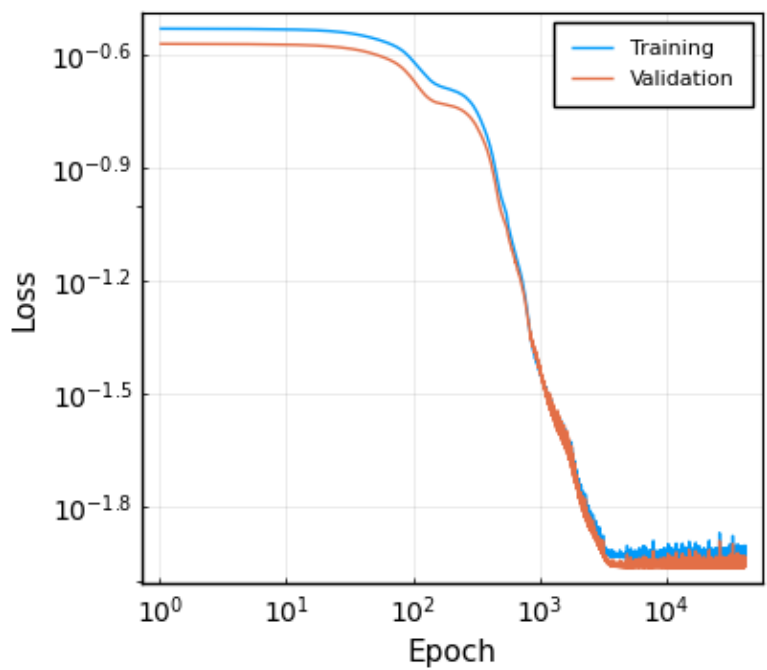

**Figure 3.** Typical evolution of loss functions with the number of epochs. Results shown correspond to the CRNN with four hidden nodes.

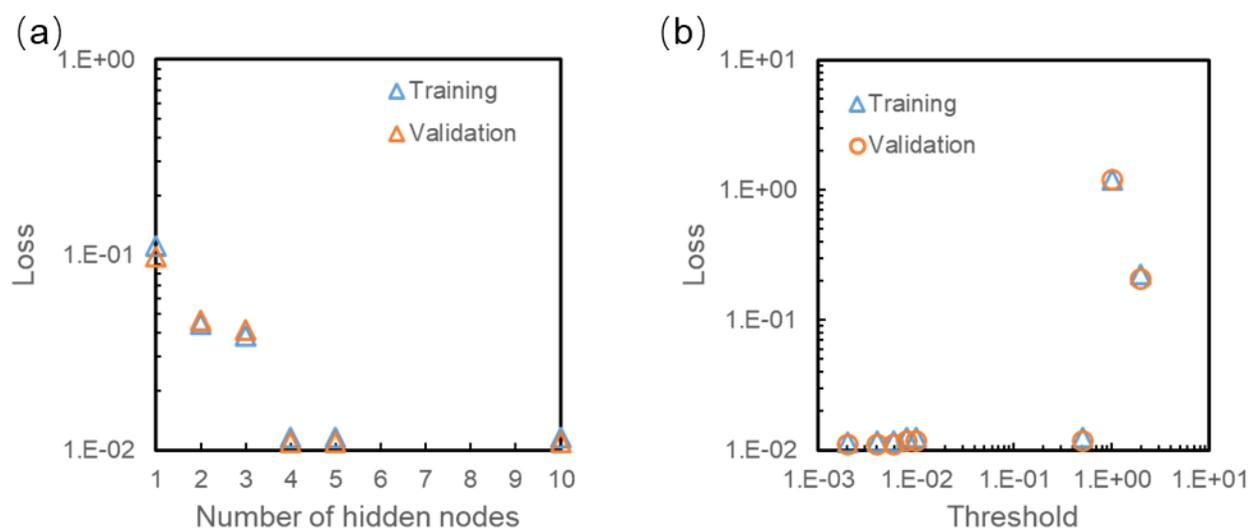

**Figure 4.** The dependence of minimum loss functions for training and validation dataset with (a) the number of hidden nodes and (b) the pruning threshold.

We then learn a CRNN with four hidden nodes and present the learned weights and biases in Fig. **5a**. The dimensions of inputs and hidden nodes are physically interpretable. Each row of the matrix corresponds to a reaction. The weights in the input layer reveal the reaction orders of each reaction, and they are always positive. The weights in the output layer correspond to the stoichiometric coefficients, and positive/negative values indicate that the corresponding species are produced/consumed, while zero values indicate that the corresponding species do not participate in the reaction. The biases reflect the rate constants. Consequently, we can infer the corresponding four reaction pathways from the learned CRNN weights.

(a) Learned CRNN weights before pruning.

|     | A     | B  | C     | D     | E     | $k$   | A      | B      | C      | D      | E      |
|-----|-------|----|-------|-------|-------|-------|--------|--------|--------|--------|--------|
| R1  | 0.    | 1  | 0.003 | 1.006 | 0.    | 0.307 | 0.002  | -1     | -0.003 | -1.006 | 1.006  |
| R2  | 2.093 | 0. | 0.    | 0.002 | 0.008 | 0.101 | -2.093 | 1.107  | 0.003  | -0.002 | -0.008 |
| R3  | 1.004 | 0. | 0.    | 0.    | 0.    | 0.206 | -1.004 | 0.005  | 0.965  | 0.004  | 0.003  |
| R4  | 0.    | 0. | 0.999 | 0.    | 0.004 | 0.13  | 0.     | 0.002  | -0.999 | 1.011  | -0.004 |

(b) Learned CRNN weights after pruning with threshold = 0.01.

|     | A     | B  | C     | D     | E  | $k$   | A      | B     | C      | D      | E     |
|-----|-------|----|-------|-------|----|-------|--------|-------|--------|--------|-------|
| R1  | 0.    | 1  | 0.    | 1.006 | 0. | 0.307 | 0.     | -1    | 0.     | -1.006 | 1.006 |
| R2  | 2.093 | 0. | 0.    | 0.    | 0. | 0.101 | -2.093 | 1.107 | 0.     | 0.     | 0.    |
| R3  | 1.004 | 0. | 0.    | 0.    | 0. | 0.206 | -1.004 | 0.    | 0.965  | 0.     | 0.    |
| R4  | 0.    | 0. | 0.999 | 0.    | 0. | 0.13  | 0.     | 0.    | -0.999 | 1.011  | 0.    |

**Figure 5.** The learned CRNN weights and biases for the case I with five inputs and four hidden nodes. (a) Learned CRNN weights before pruning. (b) Learned CRNN weights after pruning with threshold = 0.01.

There are relatively small (compared to unity) non-zero input and output weights in the learned CRNN model. To further encourage sparsity, we apply a hard threshold pruning to the input and output weights. Specifically, all the weights that have absolute values below the threshold are clipped to zero. The threshold is also determined via grid searching. Figure **4b** shows the dependence of the loss function with the threshold value. It is found that the loss function, i.e., the model performance, is insensitive to the value of the threshold when the threshold is much smaller than unity. For example, the threshold of 0.01 and 0.5 makes no difference in model performance. Figure **5b** shows the learned CRNN weights after pruning with the threshold of 0.01. The weights and bias are further translated to reaction equations and rate constants and they are shown in Table **1**. The learned pathways are close to the ground truth as the reactants and products are correctly learned. Furthermore, the learned stoichiometric coefficients and rate constants are also close to the ground truth values, with the maximum relative error within 10%. Finally, the noisy species concentration profiles and the predictions using the learned CRNN model presented in Table **1** are shown in Fig. **6**, and they agree very well.

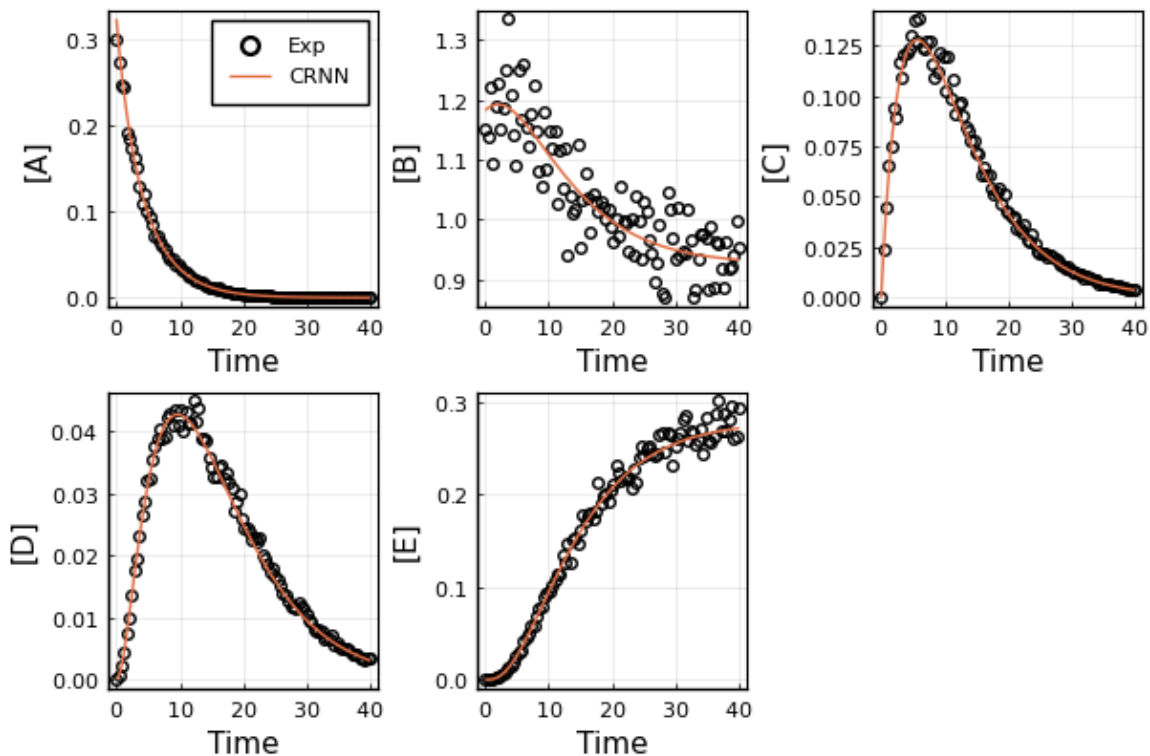

**Figure 6.** The noisy species concentration profiles and the predictions using the learned CRNN model presented in Table **1**.

It is worth mentioning that the learned stoichiometric coefficients and reaction orders are very close to integers, although we have not applied any regularization to force them to be close to integers. This makes the approach suitable for learning elementary reactions, in which the stoichiometric coefficients and reaction orders are usually assumed to be an integer. In addition, the stoichiometric coefficients and reaction orders for species not participating are learned to be very close to zero.

It is not surprising that the learned CRNN model is not exactly the same as the ground truth. The uncertainties in the chemical kinetic model parameters depend on the noise level in the data, the amount of data, and the design of initial conditions. The parameter uncertainties can be estimated using Bayesian inference [17], and the computational framework developed for traditional chemical reaction networks can be extended to the CRNN framework straightforwardly. As a rule of thumb, the model parameter uncertainties will be reduced as the experimental data uncertainties being reduced. We then present the learned CRNN model under 1% noise in Table 2. As can be seen, the learned stoichiometric coefficients are closer to the ground truth compared to the results under 5% noise. In future work, we shall employ the design of experiments [29] into the modeling pipeline to reduce the model parameter uncertainties under a moderate level of noise, e.g., 5% noise, and reduce the number of experimental datasets required.

**Table 2.** Learned pathways interpreted from the CRNN for Case I trained with noisy data with the standard deviation of the noise being 1% of the concentrations. Weights are pruned with a threshold of 0.0045.

| Ground Truth | | Learned CRNN | |
|---|---|---|---|
| Equation | Rate | Equation | Rate |
| $B + D \rightarrow E$ | 0.3 | $B + 1.002D \rightarrow E$ | 0.303 |
| $2A \rightarrow B$ | 0.1 | $2.017A \rightarrow 1.013B$ | 0.099 |
| $A \rightarrow C$ | 0.2 | $0.999A \rightarrow 0.991C$ | 0.201 |
| $C \rightarrow D$ | 0.13 | $0.999C \rightarrow 1.004D$ | 0.13 |

### 3.2 Case II: Bio-diesel Production with Temperature Dependence

The second case is to demonstrate the capability of learning the temperature dependence of the rate constants, i.e., the Arrhenius parameters in Eq. **4**. The reaction system, studied in Burnham et al. [12], is for biodiesel production via the transesterification of large, branched triglyceride (TG) molecules into smaller, straight-chain molecules of methyl esters. Darnoko and Cheryan [30] described three consecutive reactions, shown in Eq. **10**, which produce three byproducts, di-glyceride (DG), mono-glyceride (MG), and glycerol (GL). The governing equations are detailed in the Supporting Information Eq. **S2**. The pre-factor $A_0$ in the logarithmic scale for the three reactions are $[18.60, 19.13, 7.93]$, and the activation energy is $[14.54, 14.42, 6.47]$ kcal/mol.

A total of 30 experiments are simulated, and the temperature is randomly drawn from $[323 \text{ K}, 343 \text{ K}]$, according to the experimental studies in Darnoko and Cheryan [30]. The initial reactants are TG and ROH, with initial conditions randomly sampled between $[0.2, 0.2, 0, 0, 0]$ and $[2.2, 2.2, 0, 0, 0]$. Each dataset comprises of 50 data points evenly spaced in time with a time step of 1. The 30 datasets are also randomly split into training and validation datasets with a ratio of 2:1. Same as the study in Case I, we also add 5% Gaussian noise to the simulated species profiles. The training algorithms are the same as for Case I.

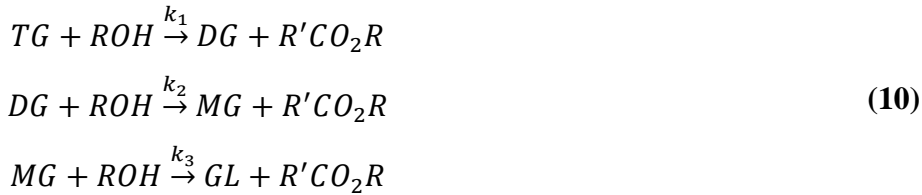

$$TG + ROH \xrightarrow{k_1} DG + R'CO_2R$$
$$DG + ROH \xrightarrow{k_2} MG + R'CO_2R \qquad (10)$$
$$MG + ROH \xrightarrow{k_3} GL + R'CO_2R$$

We also first study the dependence of the minimum loss functions with the number of hidden nodes. Typical loss curves are presented in Fig. **S1**. The training converges at around 6000 epochs. There is no obvious over-fitting observed as the training and validation loss functions are close to each other. Figure **7a** then shows that the loss functions decrease when the number of hidden nodes is less than three and almost unchanged when further increasing beyond three. Note that the model performances not necessarily strictly decrease with the number of hidden nodes due to the randomness in the training. Instead, the general trend is that the model performance becomes insensitive to the number of hidden nodes beyond three. It is then suggested that the system can be well modeled with three reactions. The dependence of the loss functions with the threshold value is further shown in Fig. **7b**. Similar to the results in Case I, the model performance is insensitive to the threshold value below unity. Figure **8** shows the weights of the learned CRNN model before pruning and after pruning with a threshold of 0.015, and the interpreted CRNN is compared with the ground truth in Table **3**. Overall, the reaction pathways are accurately learned with the maximum error in the stoichiometric coefficients within 10%. Finally, the noisy species concentration profiles and the predictions using the learned CRNN model are shown in Fig. **9**, and they agree very well. The associated model is denoted as CRNN M1 in Fig. **9**.

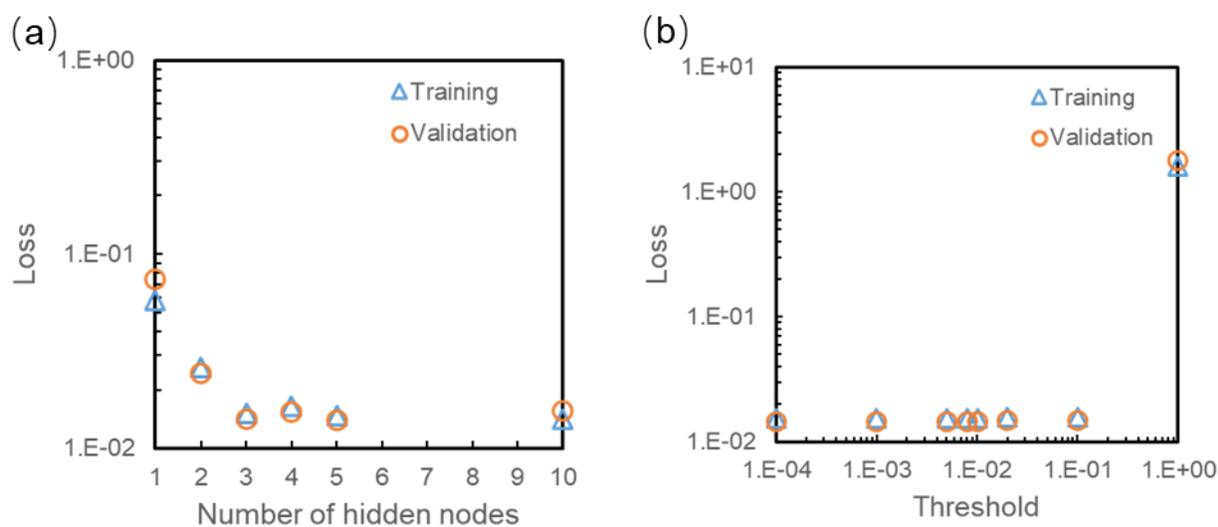

**Figure 7.** The dependence of minimum loss functions for the training and validation dataset with (a) the number of hidden nodes and (b) the pruning threshold.

(a) Learned CRNN weights before pruning.

|    | TG | ROH | DG | MG | GL | R'CO₂R | $E_a$ | $lnA$ | TG | ROH | DG | MG | GL | R'CO₂R |
|----|----|-----|----|----|----|--------|-------|-------|----|-----|----|----|----|--------|
| R1 | 1. | 0.99 | 0. | 0. | 0. | 0. | 14.44 | 18.42 | -1. | -0.99 | 1.01 | 0. | 0. | 1. |
| R2 | 0. | 0.99 | 0. | 0.99 | 0. | 0. | 6.42 | 7.81 | 0.01 | -0.99 | 0.01 | -0.99 | 1.01 | 1.09 |
| R3 | 0.01 | 1. | 1. | 0. | 0. | 0. | 14.43 | 19.18 | -0.01 | -1. | -1. | 0.97 | 0. | 0.95 |

(b) Learned CRNN weights after pruning with threshold = 0.015.

|    | TG | ROH | DG | MG | GL | R'CO₂R | $E_a$ | $lnA$ | TG | ROH | DG | MG | GL | R'CO₂R |
|----|----|-----|----|----|----|--------|-------|-------|----|-----|----|----|----|--------|
| R1 | 1. | 0.99 | 0. | 0. | 0. | 0. | 14.44 | 18.42 | -1. | -0.99 | 1.01 | 0. | 0. | 1. |
| R2 | 0. | 0.99 | 0. | 0.99 | 0. | 0. | 6.42 | 7.81 | 0. | -0.99 | 0. | -0.99 | 1.01 | 1.09 |
| R3 | 0. | 1. | 1. | 0. | 0. | 0. | 14.43 | 19.18 | 0. | -1. | -1. | 0.97 | 0. | 0.95 |

**Figure 8.** The learned CRNN weights and biases for Case II. (a) Learned CRNN weights before pruning. (b) Learned CRNN weights after pruning with a threshold of 0.015.

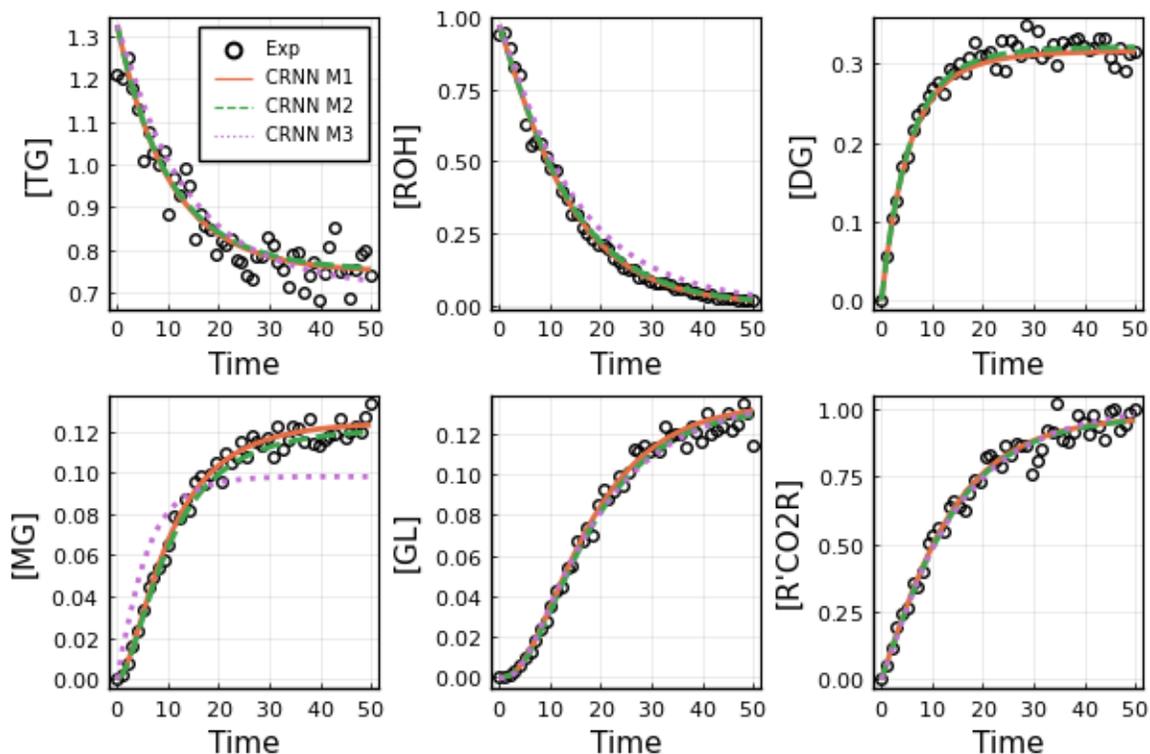

**Figure 9.** The noisy species concentration profiles and the predictions using the learned CRNN models. M1 corresponds to the CRNN learned from complete dataset, M2 corresponds to the CRNN learned from incomplete dataset with DG included in the CRNN inputs and outputs, and M3 corresponds to the CRNN learned from incomplete dataset but excluding DG from the CRNN inputs and outputs.

**Table 3.** Learned pathways interpreted from the CRNN for Case II trained with noisy data with the standard deviation of the noise being 5% of the concentrations.

| Ground Truth | | | Learned CRNN | | |
|---|---|---|---|---|---|
| Equation | $E_a$ | $lnA$ | Equation | $E_a$ | $lnA$ |
| $TG + ROH \rightarrow DG + R'CO_2R$ | 14.54 | 18.60 | $TG + 0.99ROH \rightarrow 1.01DG + R'CO_2R$ | 14.44 | 18.42 |
| $MG + ROH \rightarrow GL + R'CO_2R$ | 6.47 | 7.93 | $0.99MG + 0.99ROH \rightarrow 1.01GL + 1.09R'CO_2R$ | 6.42 | 7.81 |
| $DG + ROH \rightarrow MG + R'CO_2R$ | 14.42 | 19.13 | $DG + ROH \rightarrow 0.97MG + 0.95R'CO_2R$ | 14.43 | 19.18 |

While the above demonstrations require training the CRNN with complete datasets, assuming that the concentrations of all species are measured, such complete datasets might not always exist in real applications. We then explore the feasibility to train CRNN with incomplete datasets. Without loss of generality, we assume that the concentration history of the intermediate species DG is missing. Then we use the same data generation and training algorithms as previously discussed for training with complete datasets, except that the species DG is not included in the loss functions. Figure **S2** shows the dependence of the model performance with the number of hidden nodes, and it is suggested that the system can be modeled with three reactions. We then present the learned CRNN weights after pruning in Fig. **10**, and the predictions with the associated model denoted as CRNN M2 are shown in Fig. **9**. Satisfactorily, the learned CRNN model is also close to the ground truth, and the species profiles are well predicted. Moreover, with the learned CRNN model, we are also able to infer the species profiles of unmeasured DG.

| | $TG$ | $ROH$ | $DG$ | $MG$ | $GL$ | $R'CO_2R$ | $E_a$ | $lnA$ | $TG$ | $ROH$ | $DG$ | $MG$ | $GL$ | $R'CO_2R$ |
|---|---|---|---|---|---|---|---|---|---|---|---|---|---|---|
| R1 | 0. | 1. | 1. | 0. | 0. | 0. | 13.73 | 18.07 | 0. | -1. | -1. | 0.97 | 0. | 1.07 |
| R2 | 0. | 1. | 0. | 1.01 | 0. | 0. | 6.36 | 7.77 | 0. | -1. | 0. | -1.01 | 1.01 | 1.01 |
| R3 | 1. | 1.01 | 0. | 0. | 0. | 0. | 14.85 | 19.02 | -1. | -1.01 | 1.01 | 0. | 0. | 0.97 |

**Figure 10.** The learned CRNN weights and biases for case II trained with incomplete datasets. The weights are pruned with a threshold of 0.01.

While in the above demonstration we assume that we know the existence of the intermediate species DG in the system although we cannot measure DG, there are often cases where we do not know the existence of many intermediate species in prior. For instance, chemical kinetic modeling involves discovering not only reaction pathways but also unknown species. The CRNN approach offers a flexible framework for proposing new species by treating the number of inputs and outputs as an additional hyper-parameter and using the grid searching to determine the minimum number of required species. For example, if we exclude the species DG from the CRNN inputs and outputs, the species profiles of the rest of the five species cannot be well predicted, such as the CRNN M3 shown in Fig. **9**. Therefore, the CRNN shall include at least six species. We shall further explore the potentials of CRNN in discovering unknown species in future studies.

### 3.3 Case III: An Enzyme Reaction Network

The third case is to demonstrate the learning of catalytic (enzyme) reactions in which some species are present in both reactants and products. The mitogen-activated protein kinases (MAPK) pathway, taken from Hoffmann et al. [7], is an important regulatory mechanism for biological cells to respond to stimuli and is widely involved in proliferation, differentiation, inflammation, and apoptosis. The MAPK pathway consists of multiple stages of kinases that are either inactive or active (denoted by "*"). The activation occurs due to phosphorylation catalyzed by the kinase from the previous stage, and the dephosphorylation is catalyzed by phosphatases. When the kinase is active, it can activate subsequent kinases for the next stage. Following Hoffmann et al. [7], the MAPK pathway is modeled with three stages of kinases, MAPK, MAP2K, and MAP3K. The initial stimulus is S, and the final substrate to be activated is a transcription factor TF. The ground truth reaction network consists of activation/phosphorylation and deactivation/dephosphorylation reactions, shown in Eq. **11**. The governing equations are shown in the Supporting Information Eq. **S3**. All of the reaction rate constants are assigned to be 1.0.

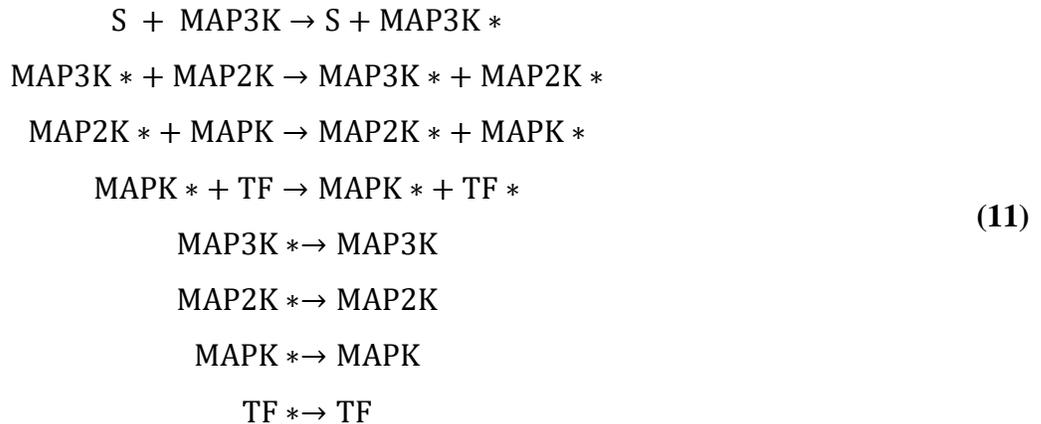

$$
\begin{aligned}
S + MAP3K &\rightarrow S + MAP3K* \\
MAP3K* + MAP2K &\rightarrow MAP3K* + MAP2K* \\
MAP2K* + MAPK &\rightarrow MAP2K* + MAPK* \\
MAPK* + TF &\rightarrow MAPK* + TF* \\
MAP3K* &\rightarrow MAP3K \\
MAP2K* &\rightarrow MAP2K \\
MAPK* &\rightarrow MAPK \\
TF* &\rightarrow TF
\end{aligned}
\tag{11}
$$

A total of 100 experiments are simulated, with initial conditions randomly sampled from the concentration of all species within [0.001, 1]. Each dataset comprises of 100 data points evenly spaced in time with a time step of 0.1. The 100 datasets are randomly split into training and validation datasets with a ratio of 70:30. Same as the study in Cases I and II, we also add 5% Gaussian noise to the simulated species profiles. The training algorithms are the same as for Case I, except that the sharing parameter between input weights and output weights are relaxed since the stoichiometric coefficients (output weights) for the catalysis could be zero while the reaction orders (input weights) are non-zero.

We also first study the dependence of minimum loss functions with the number of hidden nodes. Typical loss curves are presented in Fig. **S3**. The training converges at around 1000 epochs. Figure **11a** then shows that the loss functions decrease when the number of hidden nodes is less than eight and keep almost unchanged when further increasing the number of hidden nodes beyond eight. It is then suggested that the system can be well modeled with eight reactions. The dependence of the loss functions with the threshold value is further shown in Fig. **11b**. Similarly, the model performance is insensitive to the threshold value below unity. We then show the pruned weights in Fig. **12** with a pruning threshold of 0.01 and compared the interpreted CRNN with the ground truth in Table **4**. Overall, the reaction pathways are accurately learned with the maximum error in the stoichiometric coefficients within 3%. Finally, the noisy species concentration profiles and the predictions using the learned CRNN model are shown in Fig. **13**, and they agree very well.

Since catalytic species are present in both reactants and products, their stoichiometric coefficients shown in the output layer are zero. Together with the reaction orders inferred from the input weights, the participation of the catalyst and the catalytic pathways can be revealed, i.e., the catalyst has a positive reaction order in the input weights. The successful demonstration of using CRNN to infer catalytic reaction systems shall open the possibility of future applications in biology and chemical engineering.

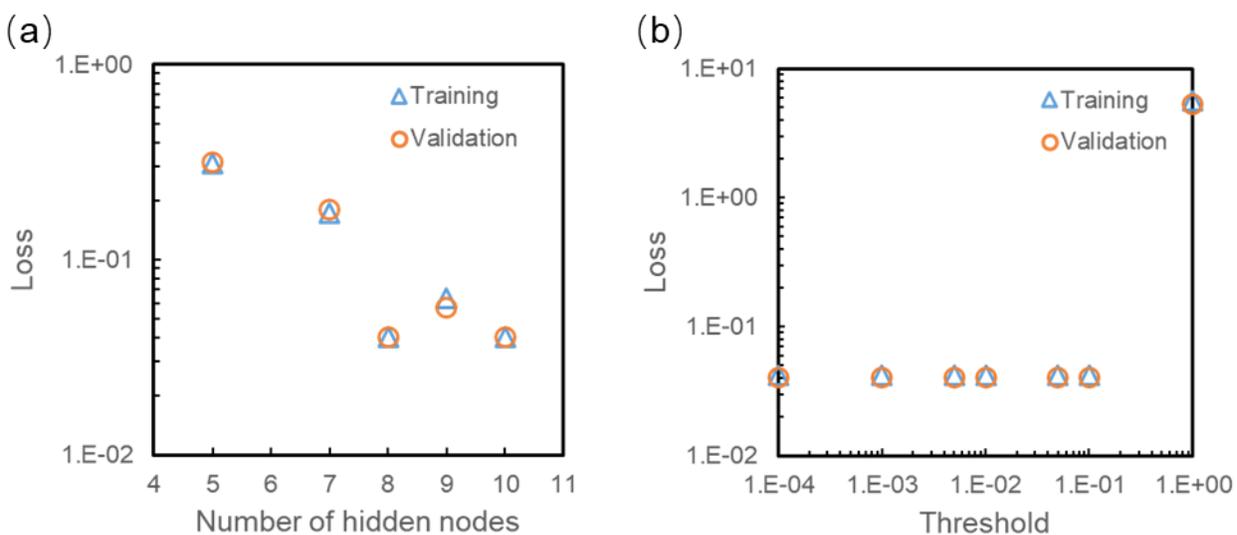

**Figure 11.** The dependence of minimum loss functions for the training and validation dataset with (a) the number of hidden nodes and (b) the pruning threshold.

| | S1 | S2 | S3 | S4 | S5 | S6 | S7 | S8 | S9 | k | S1 | S2 | S3 | S4 | S5 | S6 | S7 | S8 | S9 |
|---|---|---|---|---|---|---|---|---|---|---|---|---|---|---|---|---|---|---|---|
| R1 | 1. | 1. | 0. | 0. | 0. | 0. | 0. | 0. | 0. | 1. | 0. | -1. | 1. | 0. | 0. | 0. | 0. | 0. | 0. |
| R2 | 0. | 0. | 0. | 0. | 0. | 0. | 1. | 1.01 | 0. | 1.03 | 0. | 0. | 0. | 0. | 0. | 0. | 0. | -1. | 1. |
| R3 | 0. | 0. | 1. | 0. | 0. | 0. | 0. | 0. | 0. | 1. | 0. | 1. | -1. | 0. | 0. | 0. | 0. | 0. | 0. |
| R4 | 0. | 0. | 0. | 0. | 1. | 0. | 0. | 0. | 0. | 1. | 0. | 0. | 0. | 1. | -1. | 0. | 0. | 0. | 0. |
| R5 | 0. | 0. | 0. | 0. | 0. | 0. | 1. | 0. | 0. | 1. | 0. | 0. | 0. | 0. | 0. | 1. | -1. | 0. | 0. |
| R6 | 0. | 0. | 1. | 1. | 0. | 0. | 0. | 0. | 0. | 0.99 | 0. | 0. | 0. | -1. | 1. | 0. | 0. | 0. | 0. |
| R7 | 0. | 0. | 0. | 0. | 1. | 1. | 0. | 0. | 0. | 1. | 0. | 0. | 0. | 0. | 0. | -1. | 1. | 0. | 0. |
| R8 | 0. | 0. | 0. | 0. | 0. | 0. | 0. | 0. | 1. | 1.01 | 0. | 0. | 0. | 0. | 0. | 0. | 0. | 1. | -0.99 |

**Figure 12.** The learned CRNN weights and biases for case III. The weights are pruned with a threshold of 0.01. Indices of species: "S"(1), "MAP3K"(2), "MAP3K*"(3), "MAP2K"(4), "MAP2K*"(5), "MAPK"(6), "MAPK*"(7), "TF"(8), "TF*"(9).

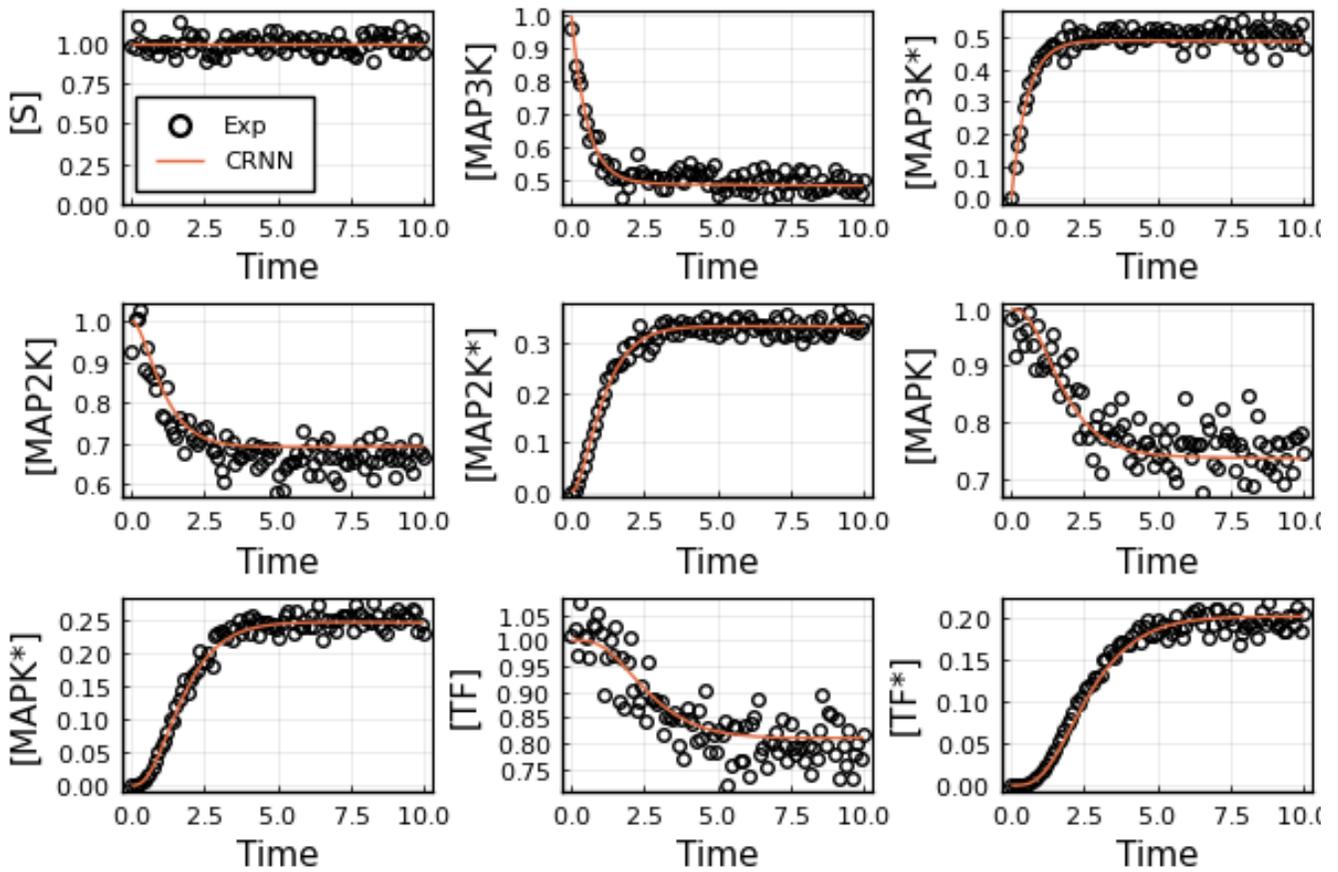

**Figure 13.** The noisy species concentration profiles and the predictions using the learned CRNN model presented in Table **4**.

**Table 4.** Learned pathways interpreted from the CRNN for Case III trained with noisy data with the standard deviation of the noise being 5% of the concentrations.

| Ground Truth | | Learned CRNN | |
|---|---|---|---|
| Equation | Rate | Equation | Rate |
| S + MAP3K → S + MAP3K ∗ | 1 | S + MAP3K → S + MAP3K ∗ | 1 |
| MAPK ∗ + TF → MAPK ∗ + TF ∗ | 1 | MAPK ∗ + TF → MAPK ∗ + TF ∗ | 1.03 |
| MAP3K ∗ → MAP3K | 1 | MAP3K ∗ → MAP3K | 1 |
| MAP2K ∗ → MAP2K | 1 | MAP2K ∗ → MAP2K | 1 |
| MAPK ∗ → MAPK | 1 | MAPK ∗ → MAPK | 1 |
| MAP3K ∗ + MAP2K → MAP3K ∗ + MAP2K ∗ | 1 | MAP3K ∗ + MAP2K → MAP3K ∗ + MAP2K ∗ | 0.99 |
| MAP2K ∗ + MAPK → MAP2K ∗ + MAPK ∗ | 1 | MAP2K ∗ + MAPK → MAP2K ∗ + MAPK ∗ | 1 |
| TF ∗ → TF | 1 | TF ∗ → 0.99 TF | 1.01 |

## 4. Discussion

Machine learning techniques, especially deep neural networks, have revolutionized the fields of computer vision [31] and natural language processing. One of the key driving forces is the stochastic gradient descent with backpropagation for high-dimension nonlinear and non-convex optimization problems, which enables the autonomous learning of features and precludes the expert knowledge in designing grammar rules. However, a common challenge for applying machine learning techniques to physical problems is that it is difficult to impose physical constraints on the model. Specifically, the neural network trained solely based on fitting training data may fail to generalize to regimes beyond the training set. In the present work, we have demonstrated an approach to encode fundamental physics laws into the neural network structure, such that the learned model satisfies the physical constraints while maintaining the accuracy in fitting the data. Therefore, the physically interpretable model is expected to generalize across a wide range of regimes and enable knowledge transfer among similar chemical systems.

Despite the successful demonstrations of the physically interpretable CRNN, further studies are needed to improve the robustness and generality of the approach. For example, the current approach assumes the chemical system is not very stiff and all of the species concentrations are within similar order of magnitude, such that we can use Eq. **8** to adequately represent the fitness of the CRNN model. While this could be a challenge when the reacting systems involve a wide range of time scales and concentration levels, approaches such as adaptive weights for the loss components [32] of different species could potentially tackle this problem by rescaling the loss functions based on the gradients of the CRNN model parameters.

Besides the encoded Law of Mass Action and Arrhenius Law, we acknowledge that the chemical systems are governed by many other physics laws as well. Encoding all of them into the structure of the neural network could be challenging. For example, the stoichiometric coefficients for elementary reactions should be integers; however, such non-differentiable constraints could be difficult to implement with stochastic gradient descent. Although we do not enforce these constraints in CRNN, the demonstrations show that the trained models automatically satisfy these constraints. Additional physical constraints such as element conservation and collision limit could potentially be included in defining the loss function to reduce the required training data and improve the generality. Identifying how "physical" the neural network needs to be is of interest for future study.

## 5. Conclusions

We have presented a Chemical Reaction Neural Network (CRNN) approach for the autonomous discovery of reaction pathways and kinetic parameters from concentration time series data. The CRNN is the digital twin of the classical chemical reaction network, and it is formulated based on the fundamental physics laws of the Law of Mass Action and Arrhenius Law. The reaction pathways and rate constants can be interpreted from the weights and biases of the CRNN. Stochastic gradient descent is adopted to optimize the large-scale nonlinear CRNN models. The approach is demonstrated in three representative chemical systems in chemical engineering and biochemistry. Both the reaction pathways and the kinetic parameters can be accurately learned. Those demonstrations shall open the possibility of discovering a large number of hidden reaction pathways in life sciences, environmental sciences, and engineering, including but not limited to gene expressions, disease progression, virus spreading, material synthesis, and energy conversion.

## 6. Associated Contents

**Supporting Information**: The governing ordinary differential equations, the loss curves and the dependence of model performance on the number of proposed reactions.

## 7. Acknowledgment

SD would like to acknowledge the support from Karl Chang (1965) Innovation Fund at Massachusetts Institute of Technology.

**Autonomous Discovery of Unknown Reaction Pathways from Data by**

**Chemical Reaction Neural Network**

Weiqi Ji, Sili Deng *

Department of Mechanical Engineering, Massachusetts Institute of Technology, Cambridge, MA 02139

Corresponding Author: * E-mail: silideng@mit.edu (Sili Deng)

ORCID: Weiqi Ji: 0000-0002-7097-0219, Sili Deng: 0000-0002-3421-7414.

## 1. Case I: An Elementary Reaction Network Without Temperature Dependence

### 1.1 1.1 Governing Equations

The governing equations for Case I are shown in Eq. **S1.**

$$\frac{d[A]}{dt} = -2k_1[A]^2 - k_2[A]$$

$$\frac{d[B]}{dt} = k_1[A]^2 - k_4[B][D]$$

$$\frac{d[C]}{dt} = k_2[A] - k_3[C] \qquad \textbf{(S1)}$$

$$\frac{d[D]}{dt} = k_3[C] - k_4[B][D]$$

$$\frac{d[E]}{dt} = k_4[B][D]$$

## 2. Case II: Bio-diesel Production with Temperature Dependence

### 2.1 2.1 Governing Equations

The governing equations for Case II are shown in Eq. **S2.**

$$\frac{d[TG]}{dt} = -r_1$$

$$\frac{d[ROH]}{dt} = -r_1 - r_2 - r_3$$

$$\frac{d[DG]}{dt} = r_1 - r_2$$

$$\frac{d[MG]}{dt} = r_2 - r_3$$

$$\frac{d[GL]}{dt} = r_3$$

$$\frac{d[R'CO_2R]}{dt} = r_1 + r_2 + r_3$$

$$r_1 = k_1[TG][ROH]$$

$$r_1 = k_2[DG][ROH]$$

$$r_1 = k_3[MG][ROH]$$

(S2)

## 2.2 2.2 Loss Curves

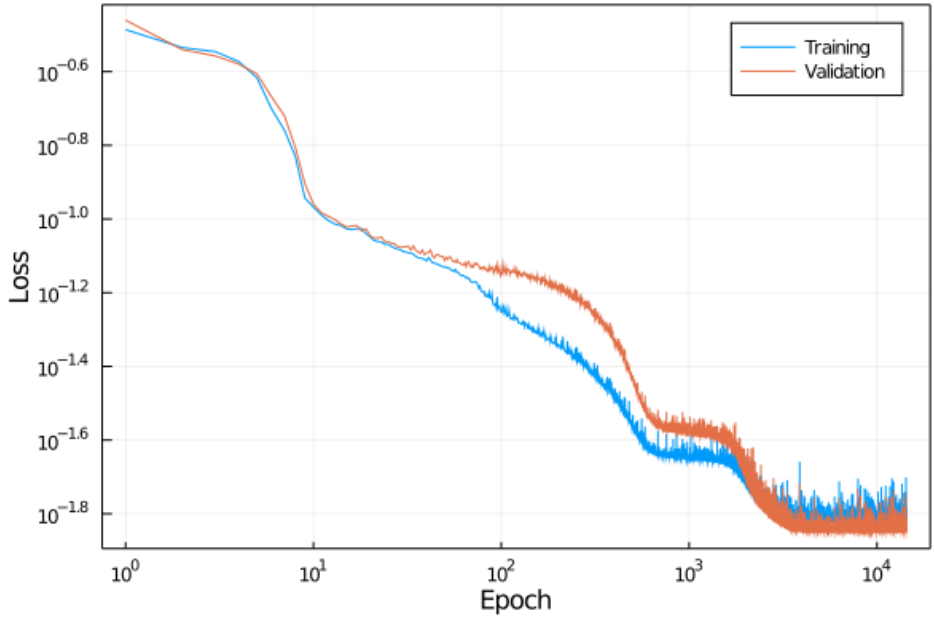

Figure. S1. The typical evolution of loss functions with the number of epochs for Case II. Results shown correspond to the CRNN with three hidden nodes.

## 2.3 2.3 Incomplete Dataset

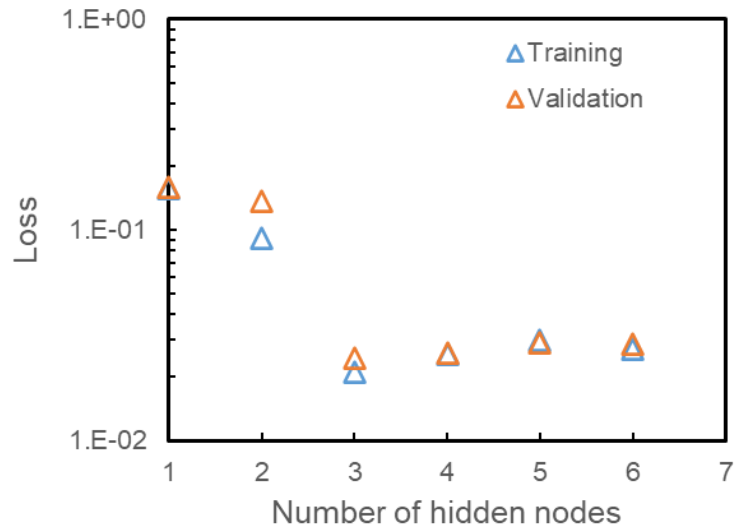

Figure. S2. The dependence of minimum loss functions for training and validation dataset with the number of hidden nodes.

## 3. Case III: An Enzyme Reaction Network

### 3.1 3.1 Governing Equations

The governing equations for Case III are shown in Eq. **S3.**

$$\frac{d[S]}{dt} = 0$$

$$\frac{d[MAP3K]}{dt} = -r_1 + r_5$$

$$\frac{d[MAP3K^*]}{dt} = r_1 - r_5$$

$$\frac{d[MAP2K]}{dt} = -r_2 + r_6$$

$$\frac{d[MAP2K^*]}{dt} = r_2 - r_6$$

$$\frac{d[MAPK]}{dt} = -r_3 + r_7$$

$$\frac{d[MAPK^*]}{dt} = r_3 - r_7$$

$$\frac{d[TF]}{dt} = -r_4 + r_8 \qquad \textbf{(S3)}$$

$$\frac{d[TF^*]}{dt} = r_4 - r_8$$

$$r_1 = k_1[S][MAP3K]$$

$$r_2 = k_2[MAP3K^*][MAP2K]$$

$$r_3 = k_3[MAP2K^*][MAPK]$$

$$r_4 = k_4[MAPK^*][TF]$$

$$r_5 = k_5[MAP3K^*]$$

$$r_6 = k_6[MAP2K^*]$$

$$r_7 = k_7[MAPK^*]$$

$$r_8 = k_8[TF^*]$$

### 3.2 3.2 Loss Curves

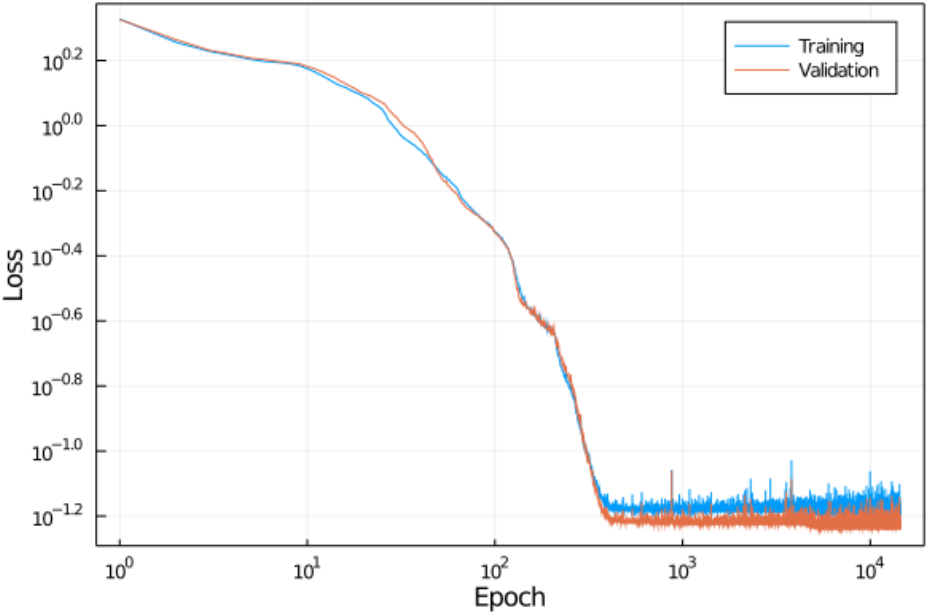

Figure S3. The typical evolution of loss functions with the number of epochs for Case III. Results shown correspond to the CRNN with eight hidden nodes.